\documentclass{pasa}

\title[Phased Array Feed Observations at Parkes]{Spectral-line Observations Using a Phased Array Feed on the Parkes Telescope}
\author[T.N. Reynolds, et al.]{T.N. Reynolds$^{1,2}$\thanks{tristan.reynolds@research.uwa.edu.au}, L. Staveley-Smith$^{1,2}$, J. Rhee$^{1,2}$, T. Westmeier$^{1}$, A.P. Chippendale$^{3}$, X. Deng$^{3,4}$, R.D. Ekers$^{3}$, M. Kramer$^4$\\
\affil{$^1$International Centre for Radio Astronomy Research (ICRAR), The University of Western Australia, 35 Stirling Hwy, Crawley, WA, 6009, Australia} \affil{$^2$ARC Centre of Excellence for All-Sky Astrophysics (CAASTRO)} \affil{$^3$CSIRO Astronomy and Space Science, Australia Telescope National Facility, P.O. Box 76, Epping NSW 1710, Australia} \affil{$^4$Max Planck Institute for Radio Astronomy, Auf dem Hügel 69, D-53121 Bonn, Germany}}

\jid{PASA}
\doi{10.1017/pas.\the\year.xxx}
\jyear{\the\year}

\usepackage[authoryear]{natbib}
\bibpunct{(}{)}{;}{a}{}{,}
\usepackage{aas_macros}
\usepackage{hyperref} 
\hypersetup{colorlinks,citecolor=blue,linkcolor=blue,urlcolor=blue}

\def\lsim{\rlap{$<$}{\lower 1.0ex\hbox{$\sim$}}}

\def\gsim{\rlap{$>$}{\lower 1.0ex\hbox{$\sim$}}}

\begin{document}

\begin{abstract}
We present first results from pilot observations using a phased array feed (PAF) mounted on the Parkes 64-m radio telescope. The observations presented here cover a frequency range from 1150 to 1480\,MHz and are used to show the ability of PAFs to suppress standing wave problems by a factor of $\sim10$ which afflict normal feeds. We also compare our results with previous HIPASS observations and with previous H\,\textsc{i} images of the Large Magellanic Cloud. Drift scan observations of the GAMA G23 field resulted in direct H\,\textsc{i} detections at $z=0.0043$ and $z=0.0055$ of HIPASS galaxies J2242-30 and J2309-30. Our new measurements generally agree with archival data in spectral shape and flux density, with small differences being due to differing beam patterns. We also detect signal in the stacked H\,\textsc{i} data of 1094 individually undetected galaxies in the GAMA G23 field in the redshift range $0.05 \leq z \leq 0.075$. Finally, we use the low standing wave ripple and wide bandwidth of the PAF to set a $3\sigma$ upper limit to any positronium recombination line emission from the Galactic Centre of $<0.09$\,K, corresponding to a recombination rate of $<3.0\times10^{45}\,\mathrm{s}^{-1}$.
\end{abstract}


\maketitle

\section{INTRODUCTION }
\label{sec:intro}

At centimetre wavelengths, the prime requirements for a sensitive radio telescope include: a large collecting area; low receiver noise; wide bandwidth; good polarisation characteristics; and immunity to radio-frequency interference (RFI). In addition, the ability to quickly survey large areas of sky requires a wide field of view, and the ability to resolve fine details requires a large diameter, or long baselines. The diversity of recent radio telescope design demonstrates that there is no unique solution to the optimum radio telescope design. For example, the Five-hundred meter Aperture Spherical radio-Telescope (FAST) \citep{Nan2011} combines a multi-feed array with a large monolithic aperture to achieve its science goals, whereas the South African MeerKAT array \citep{Jonas2009} use arrays of small dishes to achieve a good compromise between sensitivity and field of view.

However, recent developments in Phased Array Feed (PAF) technology mean that the traditional radio telescope feed, usually a large horn-like structure, is no longer the only choice of receptor. Traditional feeds can be large structures, which have low efficiencies, often have low bandwidths, and fundamentally cannot fully sample the sky at any instant. Several groups have therefore experimented with PAFs and the closely-related aperture array technologies \citep{vanArdenne2010,Roshi2015,Warnick2016}. PAFs typically consist of simple receptors closely packed on the focal plane. The voltages from these receptors are then combined in a manner that uniformly illuminates the aperture with higher efficiency than can usually be achieved by conventional means. This approach has been adopted in the CSIRO Australian SKA Pathfinder (ASKAP) telescope \citep{DeBoer2009,Hotan2014,Schinckel2016} and the ASTRON WSRT/APERTIF upgrade \citep{Oosterloo2009}.

Recently, the Max-Planck Institute for Radio Astronomy (MPIfR) procured a CSIRO-built PAF for use on the Effelsberg 100-m telescope. Prior to its installation at Effelsberg, the PAF (henceforth MPIPAF) was installed on the Parkes 64-m telescope for a 6 month commissioning workout. The MPIPAF is the Mk II version from the ASKAP Design Enhancements \citep[ADE,][]{Hampson2012} project as currently deployed on ASKAP, with some minor physical and electronic modifications designed to make it suitable for installation at Parkes and Effelsberg \citep{Chippendale2016}. In particular, the radio frequency interference (RFI) environment at these two sites is inferior to that of the ASKAP site \citep{Chippendale2013,Indermuehle2016}, which lies in the radio-quiet zone of the Murchison Radio-astronomy Observatory \citep{Wilson2013}.

The MPIPAF consists of connected dipoles in a chequerboard pattern \citep{Hay2008} and can form up to 36 dual-polarisation beams on the sky, simultaneously covering a significantly larger area in a single pointing than traditional receivers. The increased sky coverage permits larger areas of sky to be surveyed to similar sensitivity as traditional receivers, but with less observing time for cryogenically-cooled PAFs. However, one limitation to using a PAF is the increased computing requirements for forming beams and storing and processing the additional data. Only 16 beams were formed at full spectral resolution in our commissioning observations using a firmware module primarily designed for engineering verification of the ASKAP beamformer. 

Another limitation of the MPIPAF is that the measured system temperature-efficiency ratio $T_{sys}/\eta \approx 65$\,K \citep{Chippendale2016} is a factor of 1.6 times higher than the existing cryogenically-cooled 13-beam receiver at Parkes \citep{Staveley-Smith1996}. However, this is offset by the larger bandwidth and greater number of beams available. Moreover, the main purpose of the current observations was to test the viability of PAFs in large single dish reflectors and to assess their performance prior to permanently installing a cooled, high-performance version on the Parkes telescope in the future. In this work, we therefore focus on the assessment of the performance of the PAF for spectral line (mainly H\,\textsc{i}) observations. The MPIPAF has also been assessed for studying pulsars by \cite{Deng2017}.

The first part of our tests focussed on neutral hydrogen (H\,\textsc{i}) observations of the Large Magellanic Cloud (LMC). Being the closest massive gas-rich galaxy to the Milky Way Galaxy at $\sim50$\,kpc, it has been the subject of much observational study, including in H\,\textsc{i} with the Parkes 64-m telescope \citep{Mcgee1966, Bruens2005, Staveley-Smith2003}, the Australia Telescope Compact Array (ATCA) \citep{Kim2003}, and other telescopes. We use the Parkes H\,\textsc{i} survey of the LMC of \cite{Staveley-Smith2003} as a reference dataset for comparison with our MPIPAF observations.  

The second part of our tests focused on observations of extragalactic H\,\textsc{i}, which is one of the keys to understanding galaxy evolution over cosmic time. At low redshifts ($z\lesssim0.1$), we can detect and measure H\,\textsc{i} in large numbers of individual galaxies through H\,\textsc{i} spectroscopy, which involves detection of the redshifted $\lambda$21\,cm line. An example is the H\,\textsc{i} Parkes All-Sky Survey \citep[HIPASS,][]{Barnes2001} which provided a census of southern gas-rich galaxies at $z<0.04$. At higher redshifts, only the very brightest, most massive galaxies will be detected, while the average population of less massive galaxies is too faint to be detected above the telescope noise \citep{Catinella2008}. However, the H\,\textsc{i} spectra from optically identified galaxies without individual H\,\textsc{i} detections can be stacked to detect the average H\,\textsc{i} emission, as Gaussian noise fluctuations will decrease, leaving the averaged stacked H\,\textsc{i} signal \citep[e.g.:][]{Lah2007, Fabello2011, Delhaize2013, Gereb2014, Gereb2015, Brown2015, Rhee2013, Rhee2016, Kleiner2016}. For example, \cite{Delhaize2013} found stacked detections in both the South Galactic Pole and HIPASS data sets, observed on the Parkes 64-m telescope over redshift ranges $0.0405<z<0.1319$ and $z<0.0025$, respectively, using spectral stacking. Our extragalactic observations were designed to examine how well the MPIPAF was able to reproduce existing data, and whether there is any systematic noise floor that prevents detection of weak or distant spectral features.

Finally, the broad bandwidth of the MPIPAF (0.7 to 1.8 GHz) opens up two new science areas. One is `intensity mapping', which is a technique to detect the summed spectral emission from distant galaxies through their power or cross-power spectra \citep{Pen2009}. We report on these observations in a separate paper. The other is detection of recombination-line emission from positronium in the Galactic Centre. Positronium is an exotic atom composed of an electron and positron first detected in the laboratory by \cite{Canter1975}. \cite{Leventhal1978} made the first $\gamma$-ray detection of positronium annihilation from the Galactic Centre \citep[for a review of astronomical positronium studies, see][]{Ellis2009}. Radio recombination lines (RRLs) of positronium have not yet been detected with radio telescopes \citep{Anantharamaiah1989}. RRLs can be used to derive properties of diffuse gas within galaxies (e.g.: temperature and density) and are regularly observed for elements such as hydrogen, helium and carbon. The RRL frequencies of positronium can be calculated using the usual Rydberg formula.

This paper is structured as follows. We describe the MPIPAF observations on the Parkes radio telescope, and the data reduction pipeline in Section~\ref{sec:data}. In section~\ref{sec:results} we present our results including an examination of standing waves, the system temperature, and RFI in the data. We present a comparison of the data with previous observations of the LMC and individual HIPASS galaxies. We also stack H\,\textsc{i} spectra for galaxies in the GAMA G23 field and stack hydrogen and positronium recombination-line spectra in the region of the Galactic Centre. In Section~\ref{sec:conclusions} we present our conclusions. Throughout, we use J2000 coordinates, dates in UTC and adopt a flat $\Lambda$CDM cosmology using ($h$, $\Omega_{\mathrm m}$, $\Omega_{\mathrm b}$, $\Omega_\Lambda$, $\sigma_8$, $n_{\mathrm s}$) = (0.702, 0.275, 0.0446, 0.725, 0.816, 0.968), concordant with the latest \textit{WMAP} and \textit{Planck} results \citep{Bennett2013, Planck2015}.

\section{THE DATA}
\label{sec:data}

\subsection{Observations}
\label{s-sec:observations}

\begin{figure}
\centering
\includegraphics[width=\columnwidth]{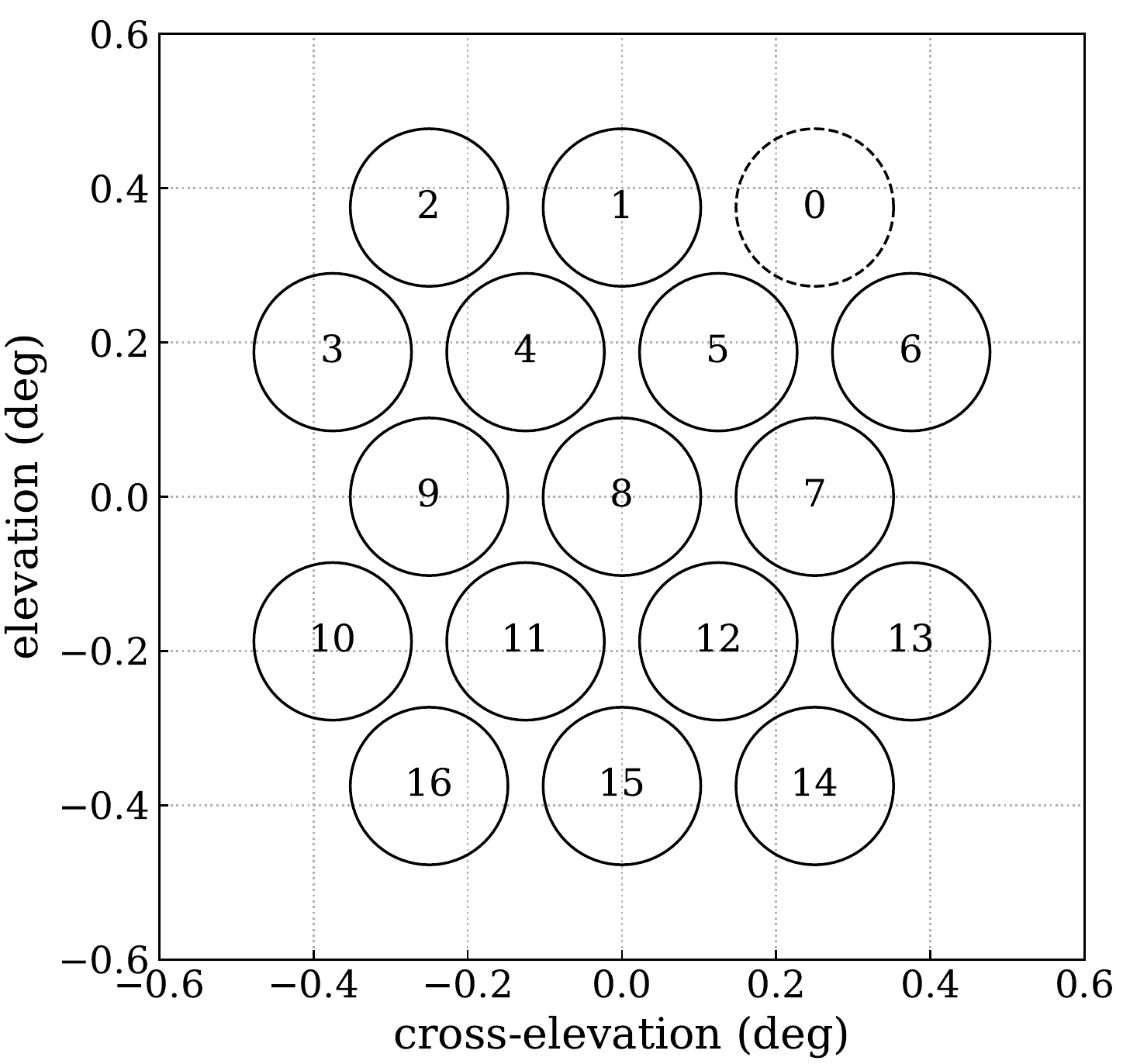}
\caption{Footprint of the MPIPAF beams with a pitch of 0.25$^{\circ}$. Only 16 of the beams were used for the observations, the beam labelled 0, dashed circle, was not used.}
\label{fig:footprint}
\end{figure}

The observations were taken using `Band 2' of the MPIPAF mounted on the Parkes 64-m radio telescope, which covers a useful band from 1200 to 1500\,MHz in two orthogonal linear polarisations. Observations were made of the LMC ($63.75^{\circ}\leq \alpha \leq  93.75^{\circ}$, $-70.5^{\circ}\leq \delta \leq -67^{\circ}$; J2000), the footprint of the GAMA survey field G23 ($339^{\circ}\leq \alpha \leq  351^{\circ}$, $-35^{\circ}\leq \delta \leq -30^{\circ}$; J2000), the Circinus galaxy, NGC\,6744 and the Galactic Centre ($\alpha,\,\delta= 17$:45:40.4,\,$-$29:00:28.1; J2000). The LMC was observed on 2016 October 6 and 24-25, the G23 field on 2016 September 2-3 and October 4-6 and 24-25, the Circinus galaxy on 2016 August 3, NGC6744 on 2016 September 1 and the Galactic Centre on 2016 September 1. The MPIPAF beamformer output 17 beams, 16 of which were able to be used for these observations (see beam footprint in Figure~\ref{fig:footprint}). The beam offsets from the central beam in the footprint were set in the beam weights using a pitch of 0.25$^{\circ}$ or 0.35$^{\circ}$. The MPIPAF specifications and observations are summarized in Table~\ref{table:paf_params} and Table~\ref{table:obs_date}, respectively. The LMC and G23 observations were taken using drift scan mode (fixed azimuth and elevation), while the Circinus galaxy, NGC\,6744 and the Galactic Centre observations were taken using on-off source pointings. Prior to each observation, the calibrator PKS\,1934-638 was also observed\footnote{Except for Circinus which used a calibrator observation from five days later (August 8), see Section~\ref{s-sec:reduction}}. We calibrated the flux density using the PKS\,1934-638 flux model from \cite{Reynolds1994}.

\begin{table}
\centering
\begin{tabular}{@{}cc@{}}
\hline
Parameter & Band 2 Values\\ \hline \hline
\vspace{-8pt} \\
Bandwidth & 384\,MHz \\
Central Frequency & 1340\,MHz \\
Spectral Resolution & 18.5\,kHz \\
Cycle Time & 4.5\,s \\
Polarisations & 2 \\
Beams & 16 \\ \hline \hline
\end{tabular}
\caption{MPIPAF and beamformer spectral-line mode specifications for this work.}
\label{table:paf_params}
\end{table}

\begin{table}
\centering
\begin{tabular}{@{}cccc@{}}
\hline
Target & Date (2016) & Scan & Integration\\
& & Type & Time\\ \hline \hline
\vspace{-8pt} \\
LMC & Oct 6, 24-25 & Drift & 7200\,s \\
G23 Field & Sept 2-3 & Drift & 2880\,s\\
 & Oct 4-6, 24-25 & & \\
Circinus & Aug 3 & ON-OFF & 90\,s \\
NGC\,6744 & Sept 1 & ON-OFF & 90\,s \\
Galactic & Sept 1 & ON-OFF & 90\,s \\
Centre & & & \\ \hline \hline
\end{tabular}
\caption{Target fields.}
\label{table:obs_date}
\end{table}

Additionally, we used archival data cubes from the first HIPASS data release \citep{Meyer2004} and archival Parkes multibeam and ATCA data of the LMC \citep{Staveley-Smith2003} for comparison with our HIPASS source and LMC observations, respectively.

We also require optical position and redshift information for potential H\,\textsc{i} sources to attempt blind H\,\textsc{i} stacking. We queried the NASA/IPAC Extragalactic Database (NED)\footnote{\url{https://ned.ipac.caltech.edu/}} to obtain position and redshift information for optically detected sources within the G23 field, which returned source redshift and positions from the 2DFGRS, GALEXASC, GALEXMSC and 2MASX surveys.

\subsection{Data Reduction}
\label{s-sec:reduction}

We reduced the data using the data reduction and gridding packages \textsc{livedata} and \textsc{gridzilla}\footnote{\textsc{livedata} and \textsc{gridzilla} are supported by the Australia Telescope National Facility and are available at
\url{http://www.atnf.csiro.au/computing/software/livedata/}.} \citep[for a description of \textsc{livedata} and \textsc{gridzilla}, see][]{Barnes2001}. Both \textsc{livedata} and \textsc{gridzilla} are designed for reducing Parkes multibeam data, which are in single dish \textsc{fits} format, \textsc{sdfits}. However, the raw MPIPAF data files are in \textsc{hdf5} format, so we first converted the \textsc{hdf5} files to \textsc{sdfits} format using the \textsc{python} package \textsc{fits2hdf} \citep{Price2015}. We also separated each of the 16 MPIPAF beams in the raw \textsc{hdf5} data files into separate \textsc{sdfits} files, as \textsc{livedata} can only conveniently handle up to 13 beams simultaneously (the number of beams in the Parkes multibeam receiver), and reduced each beam separately.

We performed bandpass correction on both the LMC and G23 field data with \textsc{livedata}. Prior to performing bandpass correction, we used PKS\,1934-638 to calibrate the flux density scale of the data. We smoothed both data sets using Hanning smoothing and performed bandpass calibration using a second-order robust polynomial and the \textsc{extended} and \textsc{compact} calibration methods for the LMC and G23 data, respectively.

After correcting the bandpass, we gridded the reduced data using \textsc{gridzilla}. We gridded the data using a weighted median, smoothed the data using a Gaussian kernel with full width half maximum (FWHM) of 6 arcmin and a cutoff radius of 13 arcmin and combined the two polarisations. Our final data cubes have a pixel scale of 4 arcmin by 4 arcmin and a spectral resolution of 18.5\,kHz.

The Galactic Centre and targeted HIPASS source on-off observations were not reduced using \textsc{livedata}. We reduced these on-off data separately using the on-source and off-source pointings,
\begin{equation}
\begin{array}{l}
\displaystyle S_{\nu}=\left(\frac{P_{\mathrm{on},\nu}}{P_{\mathrm{off},\nu}}-1\right)T_{\mathrm{sys}}(\nu),
\end{array}
\label{eq:on_off}
\end{equation}
where $T_{\mathrm{sys}}(\nu)$ is the system temperature determined from observing the calibrator PKS\,1934-638 prior to observing the science target, $P_{\mathrm{on},\nu}$ is the on-source pointing and $P_{\mathrm{off},\nu}$ is the off-source pointing. We accounted for the frequency dependence of $T_{\mathrm{sys}}$.

The one exception to this is the Circinus galaxy, which did not have a calibrator observation prior to or after the science observation on August 5. For Circinus, we used a PKS\,1934-638 observation from August 8 as the nearest calibrator observation. However, we believe that the calibration is reliable as only the central beam (beam 8) was used for this observation which was fairly stable over the August and September observations (see Section~\ref{s-sec:tsys} and Figure~\ref{fig:tsys}).

We then used \textsc{gridzilla} to grid the reduced Galactic Centre observation similarly to the LMC and G23 field data. However, we gridded each beam separately, using \textsc{gridzilla}'s weighted median statistic, \textsc{WGTMED}, for RFI suppression and used a top-hat smoothing kernel with a FWHM and cutoff radius of 12 and 6 arcmin, respectively. We did not use \textsc{gridzilla} for the targeted HIPASS sources as they lie in RFI free regions of the spectrum and simply combined the two polarisations from each beam with a \textsc{python} script.

\section{RESULTS}
\label{sec:results}

\subsection{Standing Waves}
\label{s-sec:waves}

Standing waves are introduced as a result of broadband signals entering the telescope along multiple paths and creating an interference pattern. The principal (`on-source') standing wave at Parkes is created by radiation reflecting from the feed towards the apex of the telescope. The frequency interval of this standing wave is $c/2F$, where $F$ is the focal length (Parkes $F/D=0.41$), and corresponds to 5.6\,MHz at Parkes. Reflections off other parts of the dish and feed support legs result in standing waves at other frequencies. Standing waves from off-source interference can be even more complex. As with other baseline artefacts, the effect of standing waves can be mitigated by careful calibration. However, this becomes more difficult if the phase or amplitude of the wave shifts over time \citep{Briggs1997}. Both \cite{Delhaize2013} and \cite{Kleiner2016} noted the presence of standing waves in Parkes multibeam data and further corrected by fitting and subtracting high-order polynomials from their spectra. Standing waves can even be problematic for radio interferometers \citep{Popping2008}.

These results can be understood by noting that the amplitude, $a$, of the standing wave relative to the power in the direct signal, $A$, is $a/A=2\gamma$, where $\gamma$ is the voltage ratio of the delayed and undelayed signals. So a scattered power of only 0.01\% will give rise to a standing wave amplitude ratio of 2\% for the multibeam, and a scattered power of only 0.0001\% will result in an amplitude ratio of 0.2\% for the MPIPAF. The large apparent difference in the reflection coefficient of the two receivers ($\sim100$) is partly a result of the increased efficiency of the MPIPAF. The higher efficiency of the MPIPAF leaves less energy available for multipath reflections from the feed, but this can explain at most a 1.4 times reduction in reflected power compared to the multibeam, as suggested from a comparison of the measured multibeam \citep{Staveley-Smith1996} and MPIPAF \citep{Chippendale2016} feed efficiencies of 50-64\% and 64-75\%.

However, this cannot be the only factor. We must also consider that the MPIPAF fully samples the focal plane, such that neighbouring beams, which are separated by 15 arcmin on the sky, also have the same high efficiency. On the other hand, the multibeam feed array undersamples the aperture plane, and neighbouring beams are separated by 28 arcmin, or $\sim 2$ beamwidths \citep{Staveley-Smith1996}. Therefore, the efficiency of this receiver for hypothetical beams separated by 15 arcmin is effectively zero -- i.e. most incident radiation is reflected. The low MPIPAF standing wave amplitude must therefore be a result of the low amounts of power reflected from the focal area around a given beam, and not just the power reflected from the beam itself. A more accurate analysis of the excellent standing wave performance of the MPIPAF relative to the traditional multibeam array requires a full electromagnetic simulation and diffraction analysis, and is outside the scope of this paper. 

Modulation of the primary beam pattern by standing waves on interferometers is a related problem \citep{Popping2008}. Using APERTIF on the Westerbork Synthesis Radio Telescope, \cite{Oosterloo2010} have shown that this can also been suppressed using PAFs. Aside from the lower standing wave, additional suppression can be obtained by utilising the frequency flexibility inherent in beamforming. For the MPIPAF, the beamformer weights can be adjusted in 1\,MHz sub-bands. As the primary reflection standing wave period of 5.6\,MHz is well sampled by the 1\,MHz resolution of the beamformer and the 90\,mm spatial sampling period of the MPIPAF chequerboard at the focal plane (spacing between PAF element feeds) it may be possible to suppress the standing wave even further via more advanced beamforming techniques in the future. 

\begin{figure}
\centering
\includegraphics[width=\columnwidth]{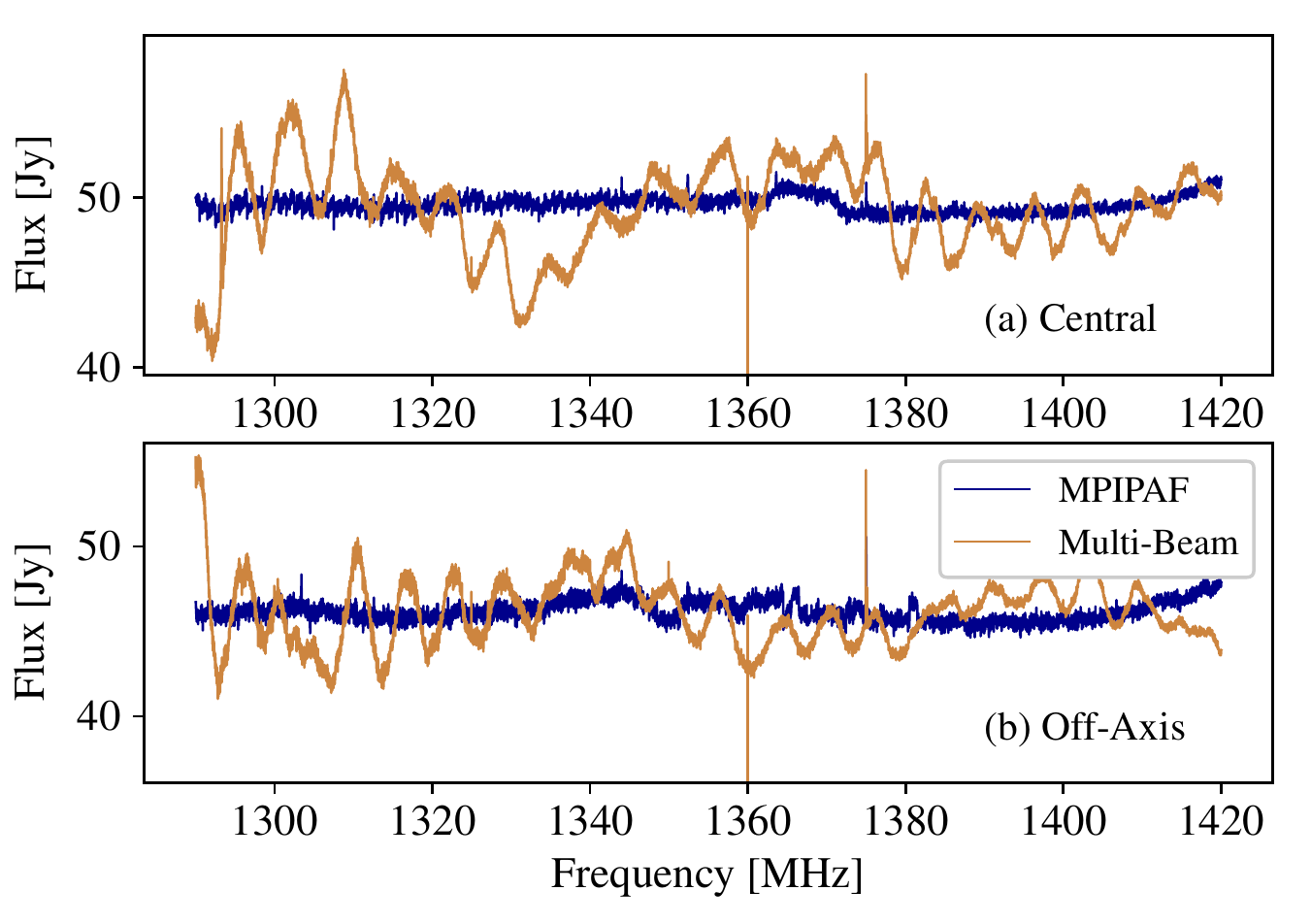}
\caption{Sample PKS\,1934-638 spectra prior to bandpass correction from the MPIPAF (blue) and multibeam (brown) central beam, panel (a), and an off-axis beam, panel (b). The angular offsets of the off-axis beams, MPIPAF beam 1 and multibeam beam 10, are $\sim0.5^{\circ}$ and $\sim1.5^{\circ}$, respectively. The MPIPAF spectra have been offset by 7 and 27\,Jy (the median difference between the MPIPAF and multibeam spectra for the central and off-axis beams, respectively) for ease of comparison with the multibeam spectra.}
\label{fig:standing_wave_spec}
\end{figure}

\begin{figure}
\centering
\includegraphics[width=\columnwidth]{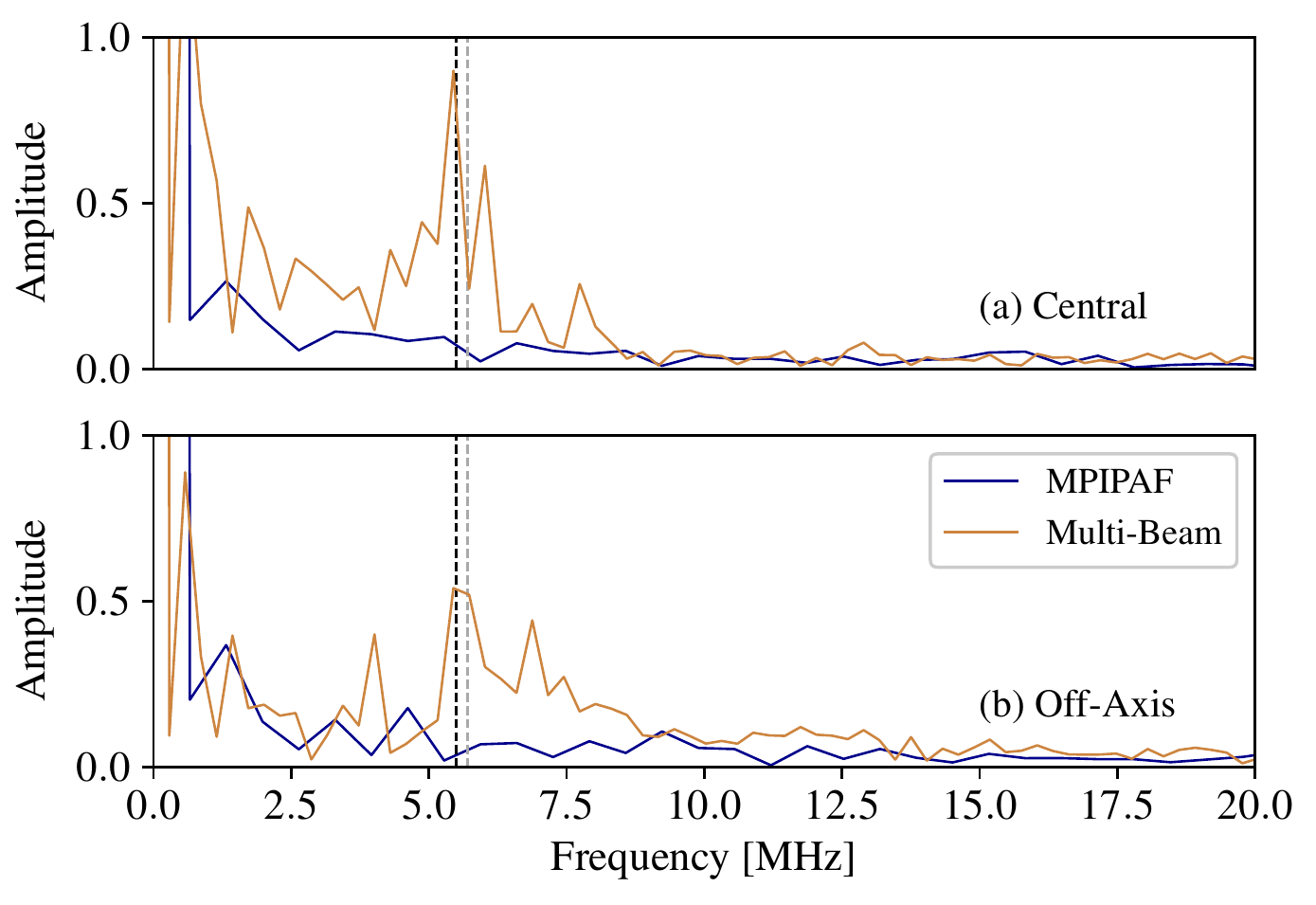}
\caption{Amplitude of the power in the standing wave spectral feature in the PKS\,1934-638 spectra in Figure~\ref{fig:standing_wave_spec} from Fourier analysis. The amplitude of the standing wave is shown for the MPIPAF (blue) and multibeam (brown) spectra in the central beam, panel (a), and an outer beam, panel (b). The vertical dashed lines indicate 5.5\,MHz and 5.7\,MHz (black and grey, respectively).}
\label{fig:standing_wave_amp}
\end{figure}

\subsection{System Temperature}
\label{s-sec:tsys}

\begin{figure}
\centering
\includegraphics[width=\columnwidth]{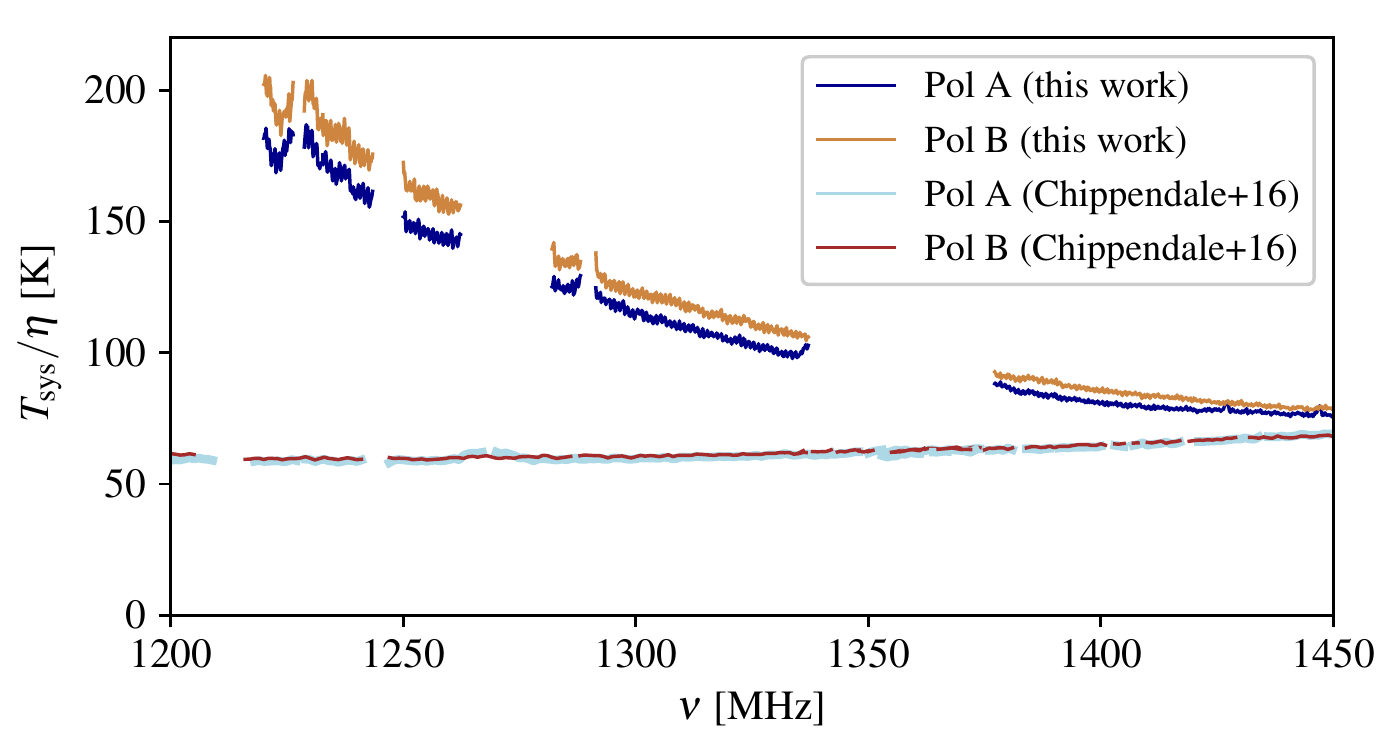}
\caption{Comparison of polarisation A and B $T_{\mathrm{sys}}/\eta$ variation with frequency for the central beam (beam 8) from October 4 observations with the test values upon installation on Parkes \citep{Chippendale2016}. The higher values measured by us at low frequencies are a result of the beamformer delay slips noted in the main text.}
\label{fig:tsys_comparison}
\end{figure}

\begin{figure}
\centering
\includegraphics[width=\columnwidth]{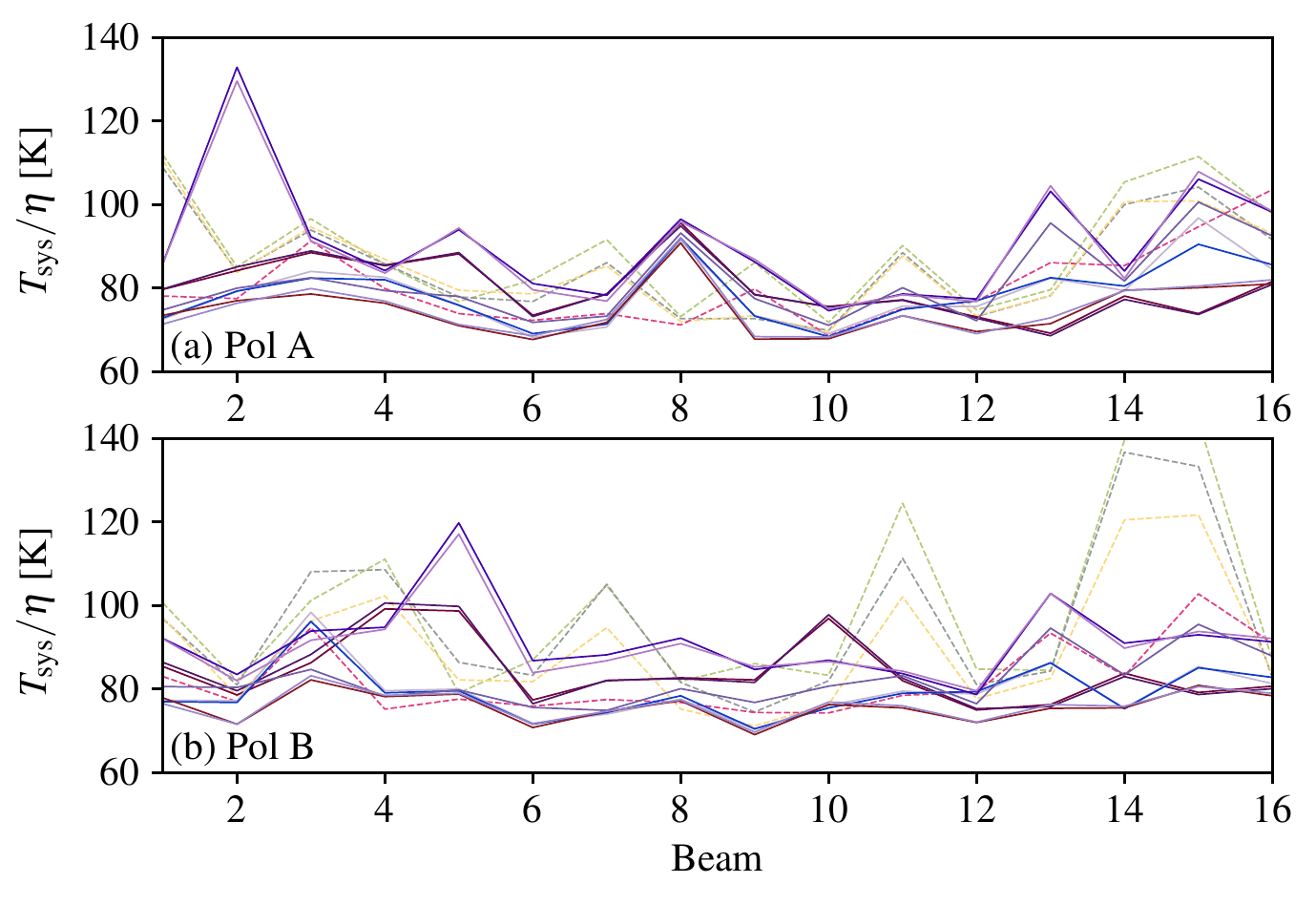}
\caption{Polarisation A and B $T_{\mathrm{sys}}/\eta$ average values over the frequency range $\nu=1400-1420$\,MHz in each of the 16 beams over observations from August, September and October (panels (a) and (b), respectively). August and September are dashed lines and October are solid lines.}
\label{fig:tsys}
\end{figure}

We determined the system temperature, $T_{\mathrm{sys}}$ using the calibrator observation of PKS\,1934-638 directly preceding the target observations. \cite{Chippendale2016} determined $T_{\mathrm{sys}}/\eta$ for the MPIPAF to be in the range $\sim45-60$\,K and to be relatively stable between the two measured polarisations from observations with the dish pointing near zenith. In the three months of observing (August-October), however, we find $T_{\mathrm{sys}}/\eta$ to be significantly higher (increasing with decreasing frequency) and less stable, ranging from $\sim70-140$\,K (mostly between$\sim70-110$\,K), with significant variation between the two polarisations and the observation dates and beams (see Figures~\ref{fig:tsys_comparison} and \ref{fig:tsys}). We also note that adjacent beams are separated by $0.25^{\circ}$, or $\sim 1$ beamwidth, so there is some overlap. We measure a mean correlation in the spectral noise between adjacent beams, in the same linear polarisation, of $\sim 10$\%. As expected, this is slightly less than the level of correlation of $13-20\%$ previously measured for adjacent ASKAP BETA beams, which are separated by $0.78^{\circ}$, or $\sim 0.7$ beamwidths \citep{Serra2015}.

Some variation in $T_{\mathrm{sys}}/\eta$ is to be expected as new beam weights were not made for each spectral line observation and the state of the hardware was not carefully controlled between observations. The probable cause of the $T_{\mathrm{sys}}/\eta$ variation are delay slips (mostly single-sample) between different ports of the MPIPAF digitiser when the digital receiver is power cycled \citep{Bannister2015}. In a production system such as ASKAP or Bonn, this can be calibrated using an on-dish noise source.

The central beam (beam 8) appears to be one of the most stable beams, remaining roughly constant during the August and September observations and at a different value for the October observations (dashed and solid lines Figure~\ref{fig:tsys}, respectively). The $T_{\mathrm{sys}}$ values for polarisation A were slightly more constant than those for polarisation B. If a PAF is permanently installed on the Parkes telescope, the $T_{\mathrm{sys}}/\eta$ noise can be lowered below the \cite{Chippendale2016} values by cooling the receiver systems. Simulations suggest that a next-generation `rocket' design can achieve $T_{\mathrm{sys}}/\eta=20-25$\,K \citep{Dunning2016}, which is a 40\,K improvement on the MPIPAF and a $15-20$\,K improvement on the Parkes multibeam receiver.

\subsection{Radio Frequency Interference}
\label{s-sec:rfi}

\begin{figure*}
\centering
\includegraphics[width=13cm,trim={0cm 8.05cm 0cm 8.25cm},clip]{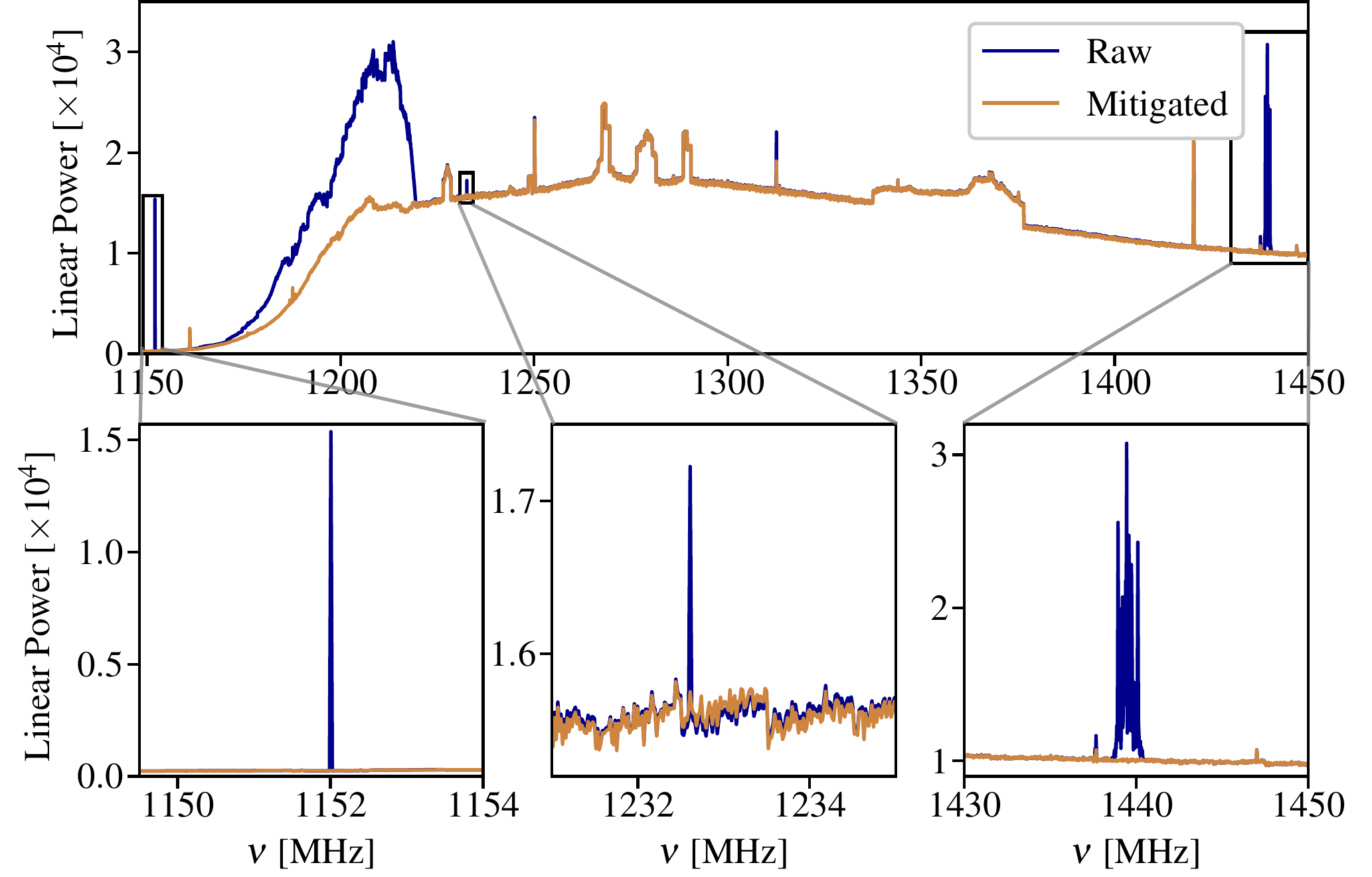}
\caption{The power spectrum (in arbitrary units) for a sample observation before (blue) and after (brown) applying a realtime projection algorithm for RFI mitigation. Substantial reduction in RFI levels is achieved for the strong RFI signals near 1150, 1210, 1230, 1310 and 1440\,MHz. We show zoomed-in spectral regions near 1150, 1230 and 1440\,MHz in the lower panels. The 1152\,MHz signal is suspected internal RFI and has also been noted in ASKAP spectra. The mitigated signal from 1159-1219.5\,MHz is most likely satellite RFI. The 1232.6\,MHz signal is an aliased fundamental of the coarse filter bank readout clock, also noted by \cite{Chippendale2017}. Based on the ACMA Register of Radiocommunications Licenses, the signal at 1439.5\,MHz is a telecommunications signal.}
\label{fig:rfimit_spec}
\end{figure*}

There is significant RFI present in the observations, as the Parkes telescope is located in New South Wales and suffers interference from radio, television, mobile phones as well as satellites.
In the frequency range of our observations, the main contributors are the navigational satellite signals (e.g. GPS) at $\nu<1290$\,MHz (see Figure~\ref{fig:rfimit_spec}). This is in contrast to the excellent RFI situation at the Murchison Radio-astronomy Observatory (MRO) where, with the exception of satellite RFI and tropospheric ducting events \citep{Indermuehle2016}, low-frequency observations are largely free of terrestrial contamination.

Since PAFs can fully sample the focal plane and have flexible beamforming capability, they are perfect for the application of advanced RFI mitigation and suppression techniques \citep[see][for an overview]{Fridman2001}. Indeed, the ASKAP Boolardy Engineering Test Array (BETA) was used to test one of these RFI mitigation methodologies, a spatial filtering technique based on projecting out the interferer signature \citep{Hellbourg2012}. This test demonstrated the effectiveness of the projection algorithm in suppressing RFI contamination in ASKAP PAF data \citep{Hellbourg2016}. This success encouraged us to attempt a similar technique for the MPIPAF data taken at the Parkes site. Full mitigation typically requires much higher temporal and spectral resolution than we have in the MPIPAF data ($\sim$\,microseconds and $\sim$\,kHz vs. 4.5\,seconds and 18.5\,kHz, respectively). However, we were able to demonstrate again the power of the projection method in some of our observations \citep[Figure~\ref{fig:rfimit_spec} shows spectra taken before (red) and after (blue) applying RFI mitigation. See][for further details]{Chippendale2017}.

For most of our observations, RFI contamination was removed from the spectra using more conventional threshold flagging techniques, by excluding individual spectral channels with flux density, $S_{\nu_{\mathrm{obs}}}>5\sigma_i$, where $\sigma_i$ is the channel RMS noise. We excluded all optical sources with redshifts placing them within the frequency range of GNSS satellites ($1240-1252$\,MHz). We also excluded, by manual inspection, all spectra containing occasional GPS L3 emissions at $1376-1384$\,MHz. We did not exclude sources falling within the receiver breakthrough at $\sim1350$\,MHz as this appeared relatively stable and, unlike other RFI regions had a reasonably low channel RMS ($\sim0.06-0.08$\,Jy, c.f. clean spectral channels RMS $\sim0.04$\,Jy).

\subsection{The Large Magellanic Cloud}
\label{s-sec:lmc}

The LMC was observed as a check of the accuracy of the flux density and frequency calibration of the MPIPAF, as well as a basic check of the reduction pipeline. We compare our results with accurate archival observations from \cite{Staveley-Smith2003}, which used the Parkes multibeam receiver. As the observed LMC emission is over the range $\nu\sim1418-1419.5$\,MHz, the LMC observations are in the RFI free section of the band and we can use all channels containing LMC emission.

\begin{figure*}
\centering
\includegraphics[width=18cm,trim={4cm 5.5cm 1cm 6cm},clip]{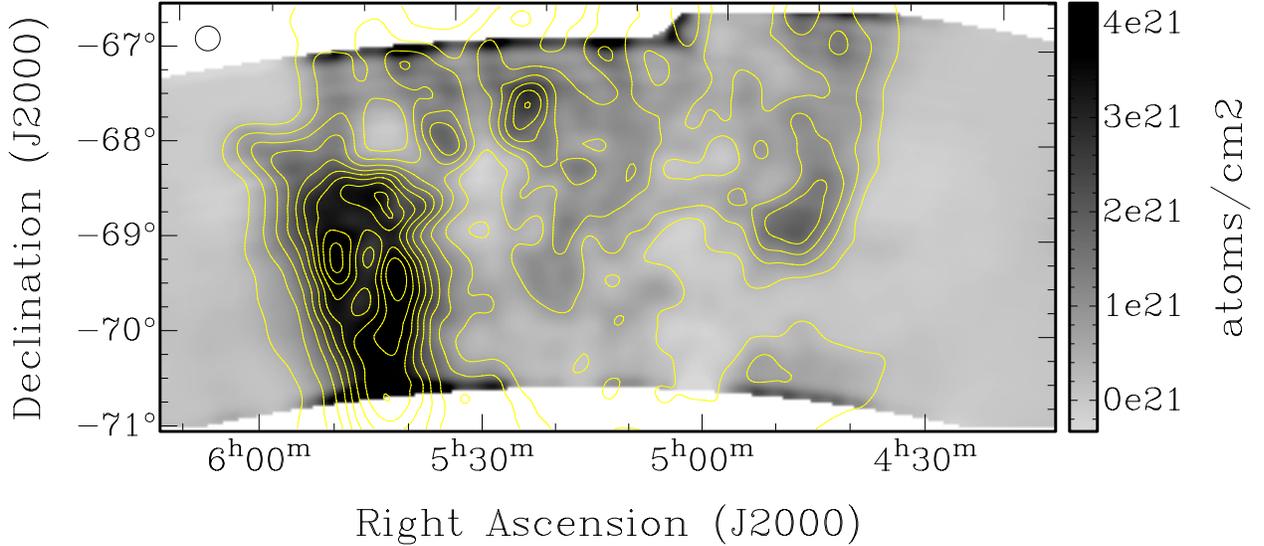}
\caption{Column density map of the MPIPAF observations of the Large Magellanic Cloud (LMC) overlaid with the yellow contours taken from the archival map from \cite{Staveley-Smith2003}. The contours are $(0.1, 0.2, 0.3, 0.4, 0.5, 0.6, 0.7, 0.8, 0.9) \times 5.58 \times 10^{21}\,\mathrm{atoms}/\mathrm{cm}^{-2}$. The beam size is shown with the black ellipse in the top left corner.}
\label{fig:lmc_mom0}
\end{figure*}

We gridded the MPIPAF data without any further calibration or adjustment except to apply a Gaussian smooth of $\mathrm{FWHM}=7$\,arcmin to remove small residual scanning lines which can still be seen faintly in Figure~\ref{fig:lmc_mom0} at $-68^{\circ}<\delta<-67^{\circ}$. Unlike the previous multibeam observations, no cross-scans were taken to mitigate against such artefacts.

In Figure~\ref{fig:lmc_mom0}, we compare the MPIPAF column density map with contours from the archival multibeam observations from \cite{Staveley-Smith2003}. The multibeam contours match well with the MPIPAF image. One thing of note in our map with the MPIPAF data are the edge effects at the top and bottom of the map which are due to the gridding process under/over-estimating the flux along the edges. 

\begin{figure}
\centering
\includegraphics[width=\columnwidth]{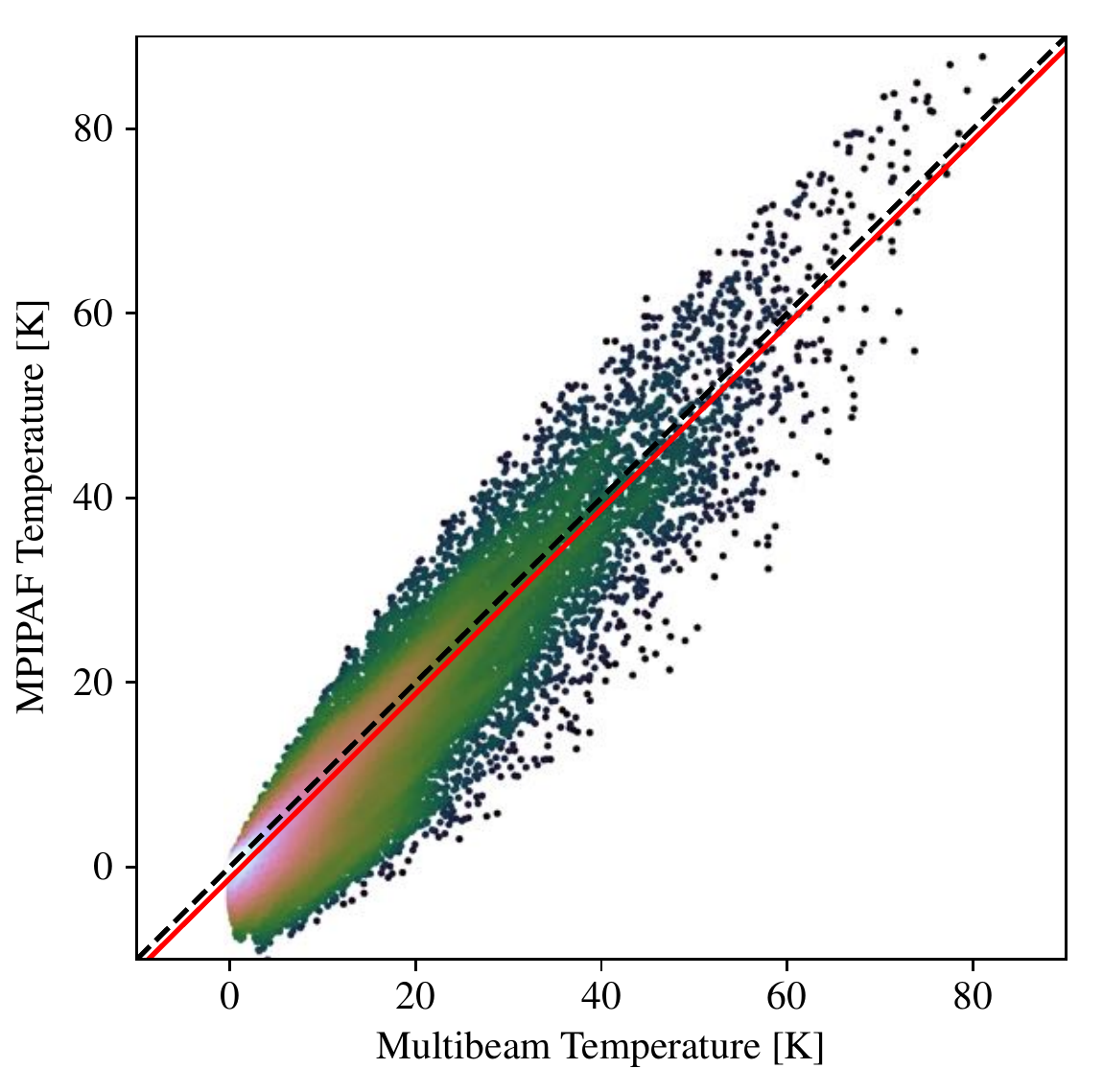}
\caption{Pixel-by-pixel comparison of temperatures in the MPIPAF and multibeam image cubes. The number of pixels compared is 96,000. The line of best fit is shown in solid red. The shading indicates the data point density, with lighter shading indicating increasing density.}
\label{fig:lmc_pixels}
\end{figure}

Figure~\ref{fig:lmc_pixels} shows a comparison of the brightness temperature values in the individual pixels of the MPIPAF and multibeam image cubes, excluding any boundary regions. There is excellent agreement, with the MPIPAF temperature having a small ($\sim1.25$\,K) zero-point offset. The zero-point offset is mainly due to the in-scan bandpass calibration procedure adopted. This has the effect of removing uniform background emission.

\subsection{GAMA G23 Field}
\label{s-sec:gama}

We extracted the spectrum for each optically identified galaxy listed in NED for the G23 field in the redshift range $0.003\leq z\leq0.23$, covering the available bandpass of the MPIPAF band 2. In each channel, we averaged the flux from a 9 pixel box ($3\times3$\,pixel $-$ $12\times12$\,arcmin) centred on the galaxy to ensure we did not lose any flux. We then extracted 600 channels ($\sim11$\,MHz) around the central redshifted frequency for each galaxy. 

\subsubsection{H\,I Stacking}
\label{ss-sec:stacking}

We perform H\,\textsc{i} stacking using the H\,\textsc{i} mass spectra rather than the originally extracted flux density spectra. We computed the observed-frame H\,\textsc{i} mass spectrum following Equation 1 from \cite{Delhaize2013},
\begin{equation}
\begin{array}{l}
\displaystyle \frac{M_{\mathrm{H\,\textsc{i},\nu_{\mathrm{obs}}}}}{M_{\odot}\,\mathrm{MHz}^{-1}}=4.98\times10^7\left(\frac{S_{\nu_{\mathrm{obs}}}}{\mathrm{Jy}}\right)\left(\frac{D_L}{\mathrm{Mpc}}\right)^2,
\end{array}
\label{eq:mass_spec}
\end{equation}
where $S_{\nu_{\mathrm{obs}}}$ is the observed-frame flux density and $D_L$ is the luminosity distance.

To stack spectra, all spectra must be shifted and aligned at the rest frequency, 1420.406\,MHz. We do this by shifting the spectral axis from observed to rest frame (i.e. $\nu_{\mathrm{rest}}=\nu_{\mathrm{obs}}(1+z)$) and to conserve total mass,
\begin{equation}
\begin{array}{l}
\displaystyle M_{\mathrm{H\,\textsc{i},\nu_{\mathrm{rest}}}}=\frac{M_{\mathrm{H\,\textsc{i}},\nu_{\mathrm{obs}}}}{1+z}.
\end{array}
\label{eq:shift_spec}
\end{equation}

We compute the stacked H\,\textsc{i} mass spectrum as done by \cite{Delhaize2013},
\begin{equation}
\begin{array}{l}
\displaystyle M_{\mathrm{stacked,i}}=\frac{\sum_i w_i M_{\mathrm{H\,\textsc{i},\nu_{\mathrm{rest}},i}}}{\sum_i w_i},
\end{array}
\label{eq:stacked_spec}
\end{equation}
where $M_{\mathrm{H\,\textsc{i},\nu_{\mathrm{rest}},i}}$ is an individual galaxy's H\,\textsc{i} mass in channel $i$, $M_{\mathrm{stacked,i}}$ is the final stacked mass in channel $i$ and $w_i$ is the weight given by
\begin{equation}
\begin{array}{l}
\displaystyle w_i=\frac{1}{\sigma_i^2 D_L^4},
\end{array}
\label{eq:weight}
\end{equation}
where $\sigma_i$ is the RMS noise in channel $i$, which we calculated using the \textsc{miriad} task \textsc{imstat}. We removed RFI contamination from the extracted spectra as described in Section~\ref{s-sec:rfi} (excluded channels with $S_{\nu_{\mathrm{obs}}}>5\sigma_i$ and excluded spectra containing GPS satellite RFI at $1376-1384$\,MHz). We then fit and subtracted a 4$^{\mathrm{th}}$-order polynomial to the stacked spectra to leave a flat baseline in the final spectrum.

We stacked the $M_{\mathrm{H\,\textsc{i}}}$ spectra in redshift bins of 0.05 up to $z=0.20$, with the final bin, $0.20 \leq z \leq 0.23$. We determined the RMS noise in the stacked spectra by randomising and reassigning each redshift to a different pair of coordinates from the input NED catalogue, ensuring that no redshift was assigned to its original coordinates. We then extracted and stacked these mock spectra identically to the galaxy spectra. The stacked mock spectra should not result in a positive detection, as our mock spectra are not centred on galaxies, and should give an approximation of the RMS noise. We performed the randomised stacking 10 times in each redshift bin and inspected each mock stacked spectrum for a possible signal mimicking a detection.

There was no detection in the $0.00 \leq z \leq 0.05$ bin after excluding two direct detections at $z=0.0043$ and $z=0.0055$ (see Section~\ref{ss-sec:detection}). Due to the increasing RFI levels at $z>0.10$, there were no direct or indirect H\,\textsc{i} detections in these redshift bins.

\begin{figure}
\centering
\includegraphics[width=\columnwidth]{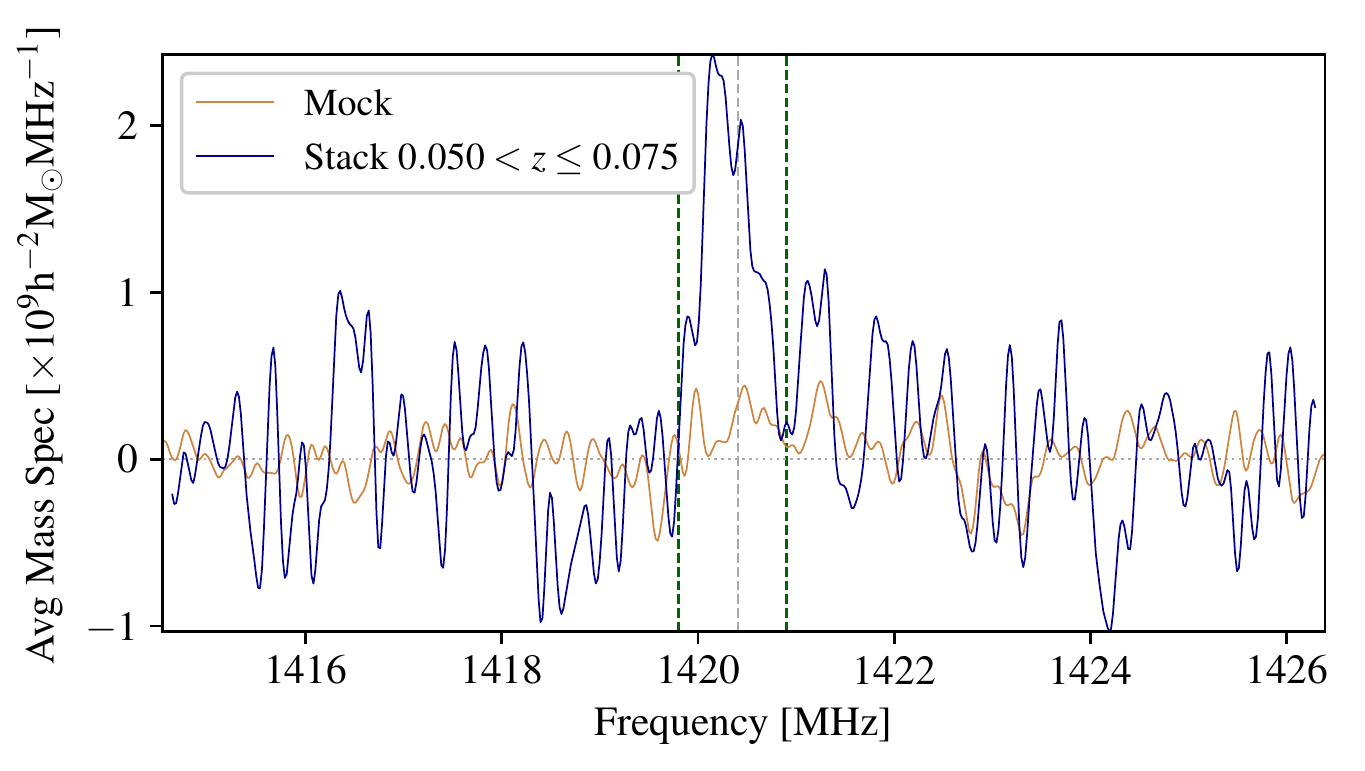}
\caption{Stacked $M_{\mathrm{H\,\textsc{i}}}$ spectrum (blue) for 1094 galaxies at $0.05 \leq z \leq 0.075$. The average mock spectrum from randomising the redshifts of the NED catalogue and stacking the spectra (shown in brown). The noise level in the mock spectrum is lower than that of the data as it is the mean of 10 simulations. The dashed green and grey lines indicate the left and right edges of the stacked H\,\textsc{i} emission determined by visual inspection and the rest frame H\,\textsc{i} line, respectively.}
\label{fig:stacked_spec}
\end{figure}

We found a detection for $\sim1100$ stacked galaxies at $0.050 \leq z \leq 0.075$. This signal was not found to be mimicked in the mock spectra from random lines of sight. In Figure~\ref{fig:stacked_spec}, we plot the stacked galaxy spectrum (blue) and the random mock spectrum (brown). Both spectra exhibit similar residual baseline curvature.

Our stacked H\,\textsc{i} detection spans the range from $\sim1419.8-1420.8$\,MHz (shown in Figure~\ref{fig:stacked_spec} by the dashed green lines), which is significantly narrower than the \cite{Delhaize2013} South Galactic Pole stacked spectra (i.e.: $\sim1$\,MHz vs. $\sim3.6$\,MHz for $z=0.05-0.075$ and $z=0.04-0.13$, respectively). Our stacked detection is most likely narrower than the results of \cite{Delhaize2013} because of our smaller sample size \citep[i.e.: $\sim1/3$ that of the South Galactic Pole region from][]{Delhaize2013} and lower redshift range, hence less confused sources entering our sample.

We integrated over the emission region to calculate the average H\,\textsc{i} mass of our stacked galaxies using
\begin{equation}
\begin{array}{l}
\displaystyle \langle M_{\mathrm{H\,\textsc{i}}}\rangle=\int^{\nu_2}_{\nu_1} \langle M_{\mathrm{H\,\textsc{i}},\nu}\rangle d\nu,
\end{array}
\label{eq:mass_avg}
\end{equation}
 where $\nu_1$ and $\nu_2$ are the edges of the emission region. We find an average integrated H\,\textsc{i} mass of $\langle M_{\mathrm{H\,\textsc{i}}} \rangle = 1.24 \pm 0.18 \times 10^9 h^{-2} M_{\odot}$. Similar to \cite{Delhaize2013}, we computed the error in $\langle M_{\mathrm{H\,\textsc{i}}} \rangle$ by integrating the $\langle M_{\mathrm{H\,\textsc{i}}} \rangle$ random mock stack. We find our $\langle M_{\mathrm{H\,\textsc{i}}} \rangle$ value to be lower than the South Galactic Pole value of $\langle M_{\mathrm{H\,\textsc{i}}} \rangle = 6.93 \pm 0.17 \times 10^9 h^{-2} M_{\odot}$ from \cite{Delhaize2013}, indicating we have detected lower mass galaxies. 

\begin{figure}
\centering
\includegraphics[width=\columnwidth]{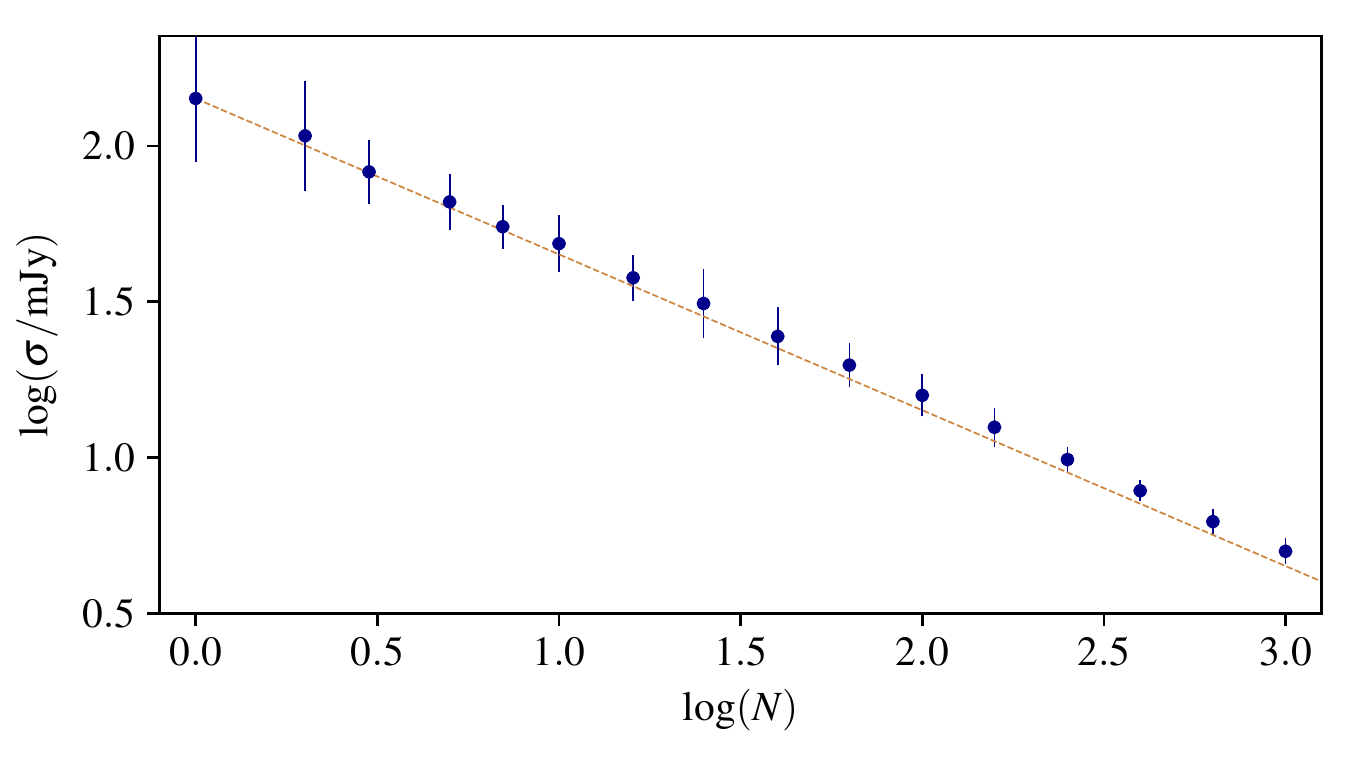}
\caption{The RMS noise in the stacked flux density signal vs. the number of stacked spectra. The error bars denote $1\sigma$ errors on the RMS. The dashed line shows the expected trend, assuming Gaussian noise, in decrease in noise with number of spectra with a gradient of $-0.5$.}
\label{fig:rms_spec_num}
\end{figure}

We investigated the noise behaviour of the stacked MPIPAF data by stacking the mock random line of sight flux density spectra, as described previously. We calculated the RMS of a stack of $N$ randomly chosen spectra to determine the noise behaviour with increasing number of stacked spectra. To estimate the noise behaviour we fix the weighting of each spectra using the mean RMS calculated from 1000 individual mock spectra ($\sigma=0.142$\,Jy) over the frequency range $1321-1352$\,MHz (corresponding to the redshift range of the stacked H\,\textsc{i} detection). Figure~\ref{fig:rms_spec_num} shows the change in RMS noise in the stacked spectrum with number of spectra included in the stack with the error bars calculated as the $1\sigma$ standard deviation of the RMS for 100 random stacks of $N$ spectra. We find the noise decreases as expected for Gaussian noise with a gradient of $-0.49 \pm 0.01$ for the MPIPAF data. This is similar to the noise behaviour for the Parkes multibeam data (gradient $\sim-0.5$) from \cite{Delhaize2013}.

\subsection{Comparison with HIPASS Detections}
\label{s-sec:hipass}

We present results of targeted observations of two HIPASS sources, J1413-65 (Circinus) and J1909-63A (NGC\,6744) from August 3 and September 4, respectively, observed with the MPIPAF, in addition to two HIPASS sources we detected in the GAMA G23 field, J2242-30 (NGC\,7361) and J2309-30 (ESO\,469-G015). We obtain HIPASS spectra from the archival HIPASS data cubes from the first HIPASS data release \citep{Meyer2004}.

\subsubsection{Targeted HIPASS Observations}
\label{ss-sec:target}

\begin{figure}
\centering
\includegraphics[width=\columnwidth]{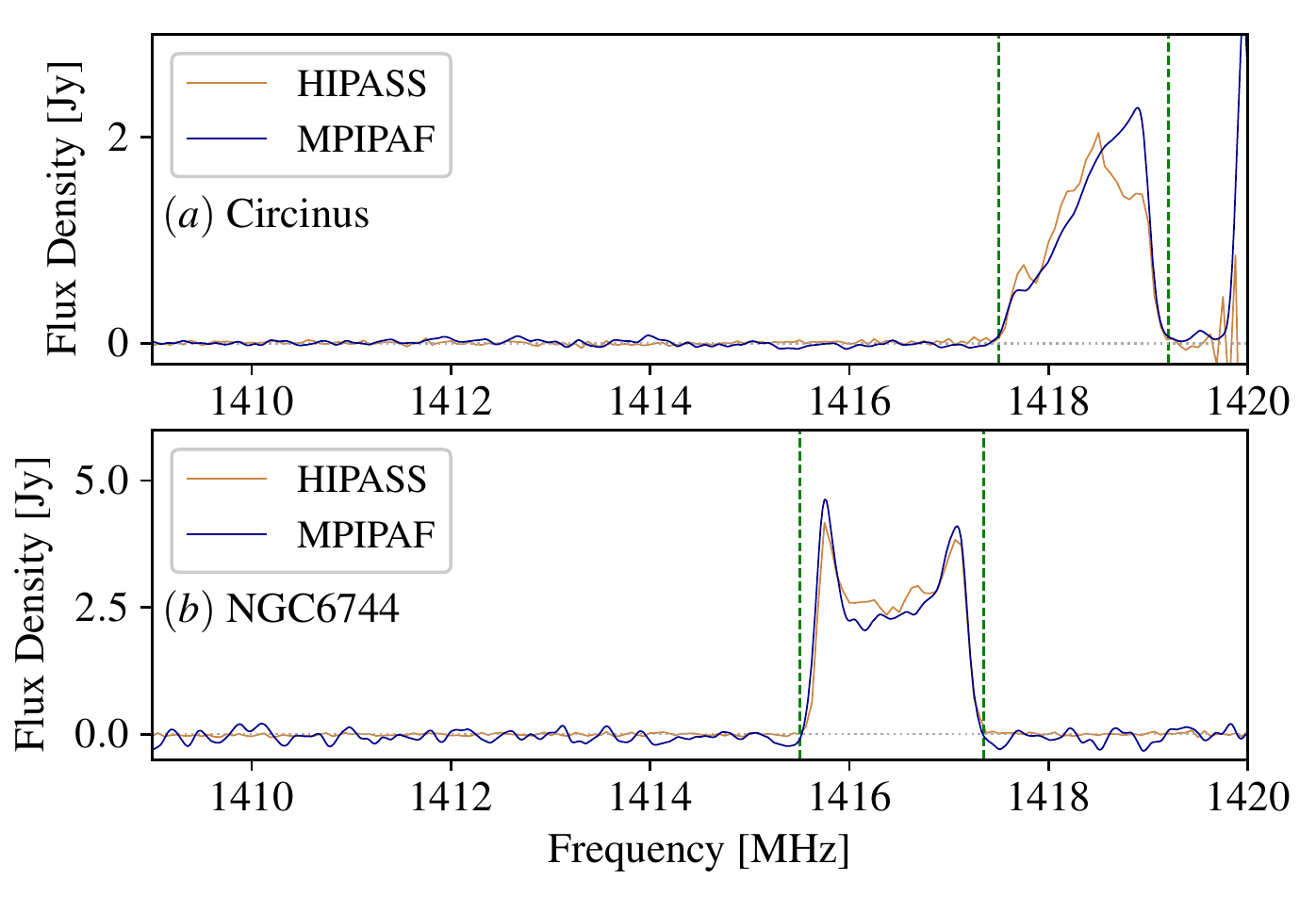}
\caption{Targeted HIPASS galaxy line of sight spectra of Circinus and NGC\,6744, panels (a) and (b), respectively. The Circinus spectrum is from a single line of sight, while the NGC 6744 spectrum is the integrated line of sight spectrum from the 16 MPIPAF beams. The MPIPAF and HIPASS spectra are shown in blue and brown, respectively. The dashed green lines indicate the edges of the galaxy emission determine by visual inspection.}
\label{fig:hipass_specs}
\end{figure}

The Circinus galaxy, was only observed with the central MPIPAF beam providing a single spectrum. We therefore compare the MPIPAF observation with the pencil beam spectrum along the same line of sight through the galaxy from the archival HIPASS data cube, as this galaxy is resolved and not contained within a single beam. The MPIPAF and HIPASS spectra agree well in shape and flux density with the only difference in that we detect some additional flux on the high frequency end of the spectrum ($\nu \gsim 1418.5$\,MHz, top panel of Figure~\ref{fig:hipass_specs}). This is most likely due to a slight position offset between the nearest HIPASS data cube pixel to the MPIPAF spectrum position (HIPASS pixel position: $\alpha,\,\delta=213.41^{\circ},\,-65.35^{\circ}$, MPIPAF line of sight position: $\alpha,\,\delta=213.36^{\circ},\,-65.31^{\circ}$).

Although the NGC\,6744 observation utilized all 16 MPIPAF beams, not all 16 beams lie upon the galaxy. NGC\,6744 has an angular radial size in H\,\textsc{i} of $\sim15$\,arcmin \citep[$\sim30$\,arcmin in diameter,][]{Ryder1999}, while 13 of the MPIPAF beams have angular separations $>30$\,arcmin from the centre of NGC\,6744. We integrated the flux from the three beams with angular separations $<24$\,arcmin. We also calculated the total integrated flux density from the archival HIPASS cube within a $13\times13$ pixel box centred on NGC\,6744 (maximum angular separation 24\,arcmin, matching the separation of the MPIPAF beams). The spectral shape and flux density of the integrated MPIPAF spectrum agrees with the HIPASS spectrum (lower panel of Figure~\ref{fig:hipass_specs}). 

\subsubsection{G23 H\,I Detections}
\label{ss-sec:detection}

\begin{figure}
\centering
\includegraphics[width=\columnwidth]{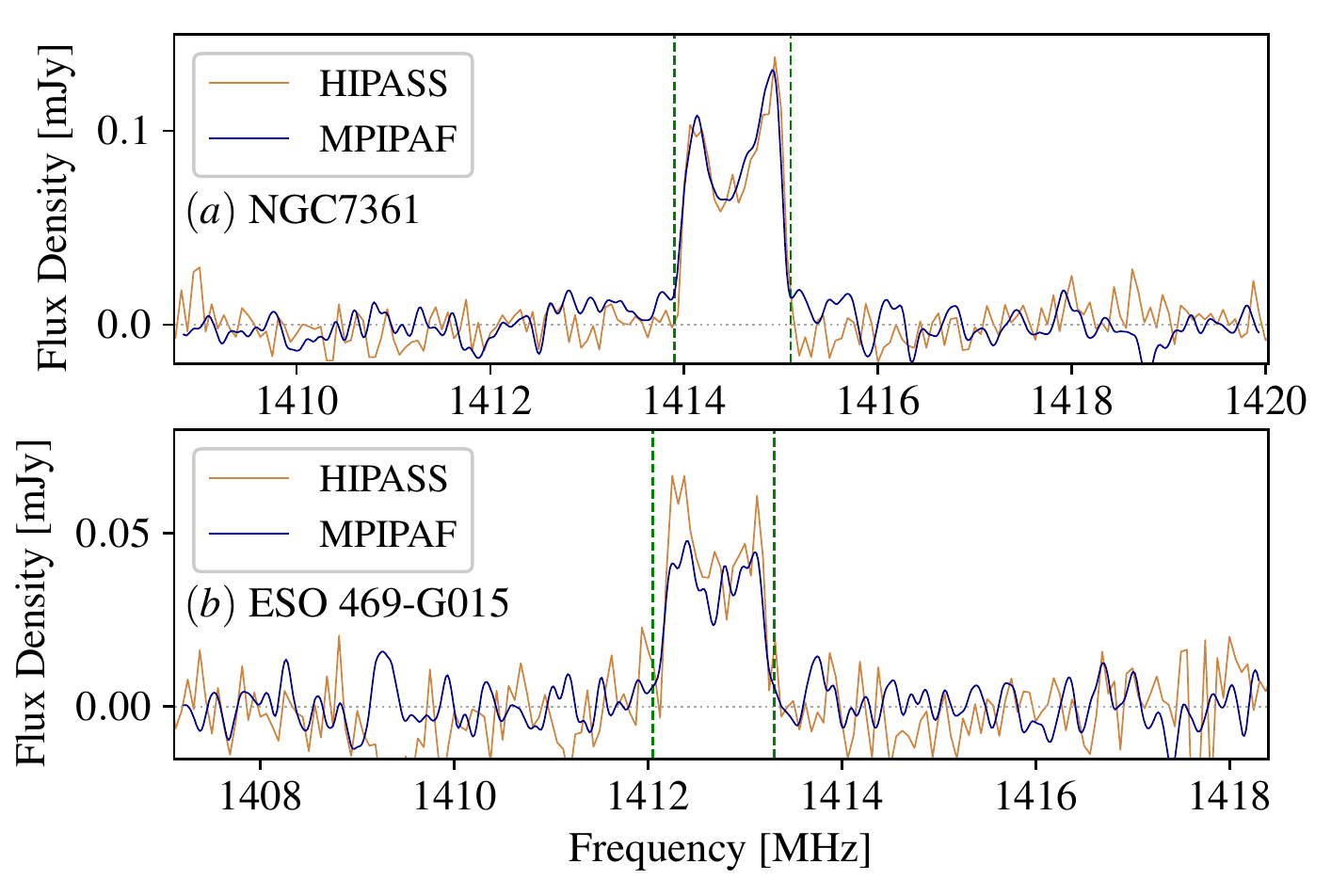}
\caption{Direct H\,\textsc{i} detection of HIPASS galaxies NGC\,7361 and ESO\,469-G015 at $z=0.0043$ and $z=0.0055$, panels (a) and (b), respectively. The MPIPAF and HIPASS integrated spectra are shown in blue and brown, respectively. The dashed green lines indicate the edges of the galaxy emission determine by visual inspection.}
\label{fig:g23_specs}
\end{figure}

We have two direct H\,\textsc{i} detections of HIPASS detected galaxies, NGC\,7361 and ESO\,469-G015, at $z=0.0043$ and $z=0.0055$ (Figure~\ref{fig:g23_specs} panels (a) and (b), respectively). We found these direct detections through visual inspection of the H\,\textsc{i} spectra extracted from the G23 field based on optical identifications from NED. Unlike the targeted HIPASS galaxy observations, we can compare the total integrated flux for these two galaxies as they are completed covered by the drift scan. We compare our MPIPAF spectra with the HIPASS spectra integrated over the same area from the archival data cubes ($20\times20$\,arcmin) for these two galaxies in Figure~\ref{fig:g23_specs} (blue and brown lines, respectively). The integrated MPIPAF spectrum of NGC 7361 shows very good agreement with the HIPASS data both in spectral shape and flux density. While the integrated MPIPAF spectrum of ESO 469-G015 has a similar spectral shape, it has a slightly lower flux density. Nevertheless, both spectra agree within the combined uncertainties (Table~\ref{table:g23_fluxes}).

\begin{table}
\centering
\begin{tabular}{@{}ccc@{}}
\hline
Galaxy & MPIPAF & HIPASS \\
 & Flux [mJy] & Flux [mJy] \\ \hline \hline
\vspace{-8pt} \\
NGC\,7361 & $0.098\pm0.014$ &$0.097\pm0.009$ \\
ESO\,469-G015 & $0.041\pm0.012$ & $0.049\pm0.011$ \\ \hline \hline
\end{tabular}
\caption{Integrated fluxes for MPIPAF and HIPASS galaxy spectra shown in Figure~\ref{fig:g23_specs}.}
\label{table:g23_fluxes}
\end{table}

\subsection{Galactic Centre Hydrogen and Positronium Recombination Lines}
\label{s-sec:galcent}

\begin{table}
\centering
\begin{tabular}{@{}cccc@{}}
\hline
H$n\alpha$ & $\nu_{\mathrm{H}}$ [MHz] & Ps$n\alpha$ & $\nu_{\mathrm{Ps}}$ [MHz] \\ \hline \hline
\vspace{-8pt} \\
165 & 1450.58 & 131 & 1446.81 \\
166 & 1424.60 & 132 & 1414.29 \\
167 & 1399.24 & 133 & 1382.75 \\
168 & 1374.48 & 134 & 1352.14 \\
169 & 1350.29 & 135 & 1322.42 \\
170 & 1326.67 & 136 & 1293.57 \\
171 & 1303.60 & 137 & 1265.55 \\
172 & 1281.06 & 138 & 1238.33 \\
173 & 1259.03 & 139 & 1211.89 \\
174 & 1237.51 & 140 & 1186.20 \\
175 & 1216.48 & 141 & 1161.23 \\
176 & 1195.92 & & \\
177 & 1175.82 & & \\
178 & 1156.17 & & \\ \hline \hline
\end{tabular}
\caption{Rydberg numbers ($n$) and corresponding hydrogen (H) and positronium (Ps) frequencies.}
\label{table:rydberg}
\end{table}

\begin{figure}
\centering
\includegraphics[width=\columnwidth]{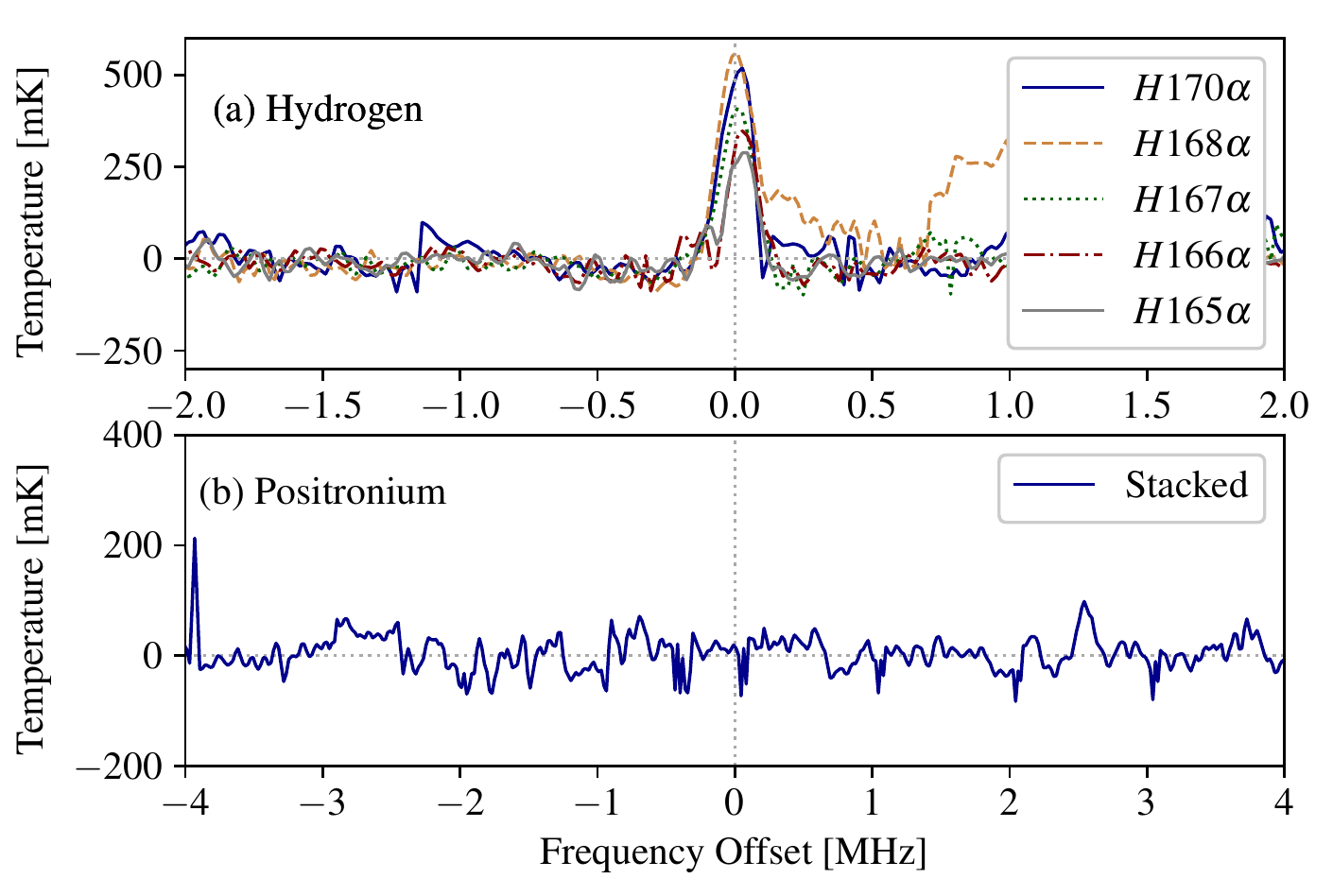}
\caption{Stacked Galactic Centre hydrogen and positronium spectra, Panels (a) and (b), respectively. For hydrogen we stacked the spectra from all 16 MPIPAF beams, for individual recombination lines and recovered a detection for the $165\alpha$, H$166\alpha$, H$167\alpha$, H$168\alpha$ and H$170\alpha$ lines. The positronium spectrum is the combined stack of the Ps$131\alpha$, Ps$132\alpha$, Ps$133\alpha$ and Ps$135\alpha$ lines in all 16 MPIPAF beams and does not show a detection.}
\label{fig:gc_stack}
\end{figure}

We inspected the Galactic Centre spectra from each beam at the 14 Hydrogen and 11 positronium radio recombination line frequencies predicted from the Rydberg equation within the band (Table~\ref{table:rydberg}). We were unable to use large sections of the spectra due to RFI contamination. Also present in the spectra are 1\,MHz beamformer `jumps', some of which were not removed during bandpass calibration. These 1\,MHz `jumps' have also been seen in early ASKAP data and are not caused by RFI or the Parkes dish, but are due to the discretization of the beamformer weights. We were able to model and remove the spectral shape of the `jumps' with a first order polynomial fit to each 1\,MHz spectral interval as the `jumps' occur at exactly 1\,MHz intervals (e.g.: 1380.5\,MHz, 1381.5\,MHz, 1382.5\,MHz, etc.). We find detections for the H$165\alpha$, H$166\alpha$, H$167\alpha$, H$168\alpha$ and H$170\alpha$ hydrogen recombination lines in individual MPIPAF beams (Figure~\ref{fig:gc_stack}(a) shows the stacked signal for these five lines in all 16 MPIPAF beams). In Table~\ref{table:h_recomb} we list calculated line parameters from Gaussian fits to the spectra in Figure~\ref{fig:gc_stack}(a). Our calculated peak intensity line temperatures agree with previous single dish Galactic Centre hydrogen RRL studies (Table~\ref{table:h_recomb}). All other hydrogen recombination lines lie within regions with high RFI contamination and are undetected. 

For positronium, however, we did not have any clear direct detections above the noise. In an attempt to improve the signal to noise, we stacked the Galactic Centre spectra from each beam centred on the predicted frequencies of positronium recombination lines to look for a stacked detection. We were unable to stack spectra at all predicted frequencies due to the presence of RFI, as mentioned above. We excluded the RFI contaminated sections of the spectra. This reduced the number of stacked positronium spectra to 64, centred on the predicted Ps$131\alpha$, Ps$132\alpha$, Ps$133\alpha$ and Ps$135\alpha$ lines. For each spectrum, we extracted 450 channels centred on the predicted line frequency and fit a $2^{\mathrm{nd}}$-order polynomial to remove the baseline from the stacked spectrum. We find no detection in the stacked positronium spectra in either emission or absorption as shown in Figure~\ref{fig:gc_stack}(b) and set a $3\sigma$ upper limit on the stacked recombination line signal of $<0.09$\,K. Using this upper limit, we calculate the recombination rate to be $<3.0\times10^{45}\,\mathrm{s}^{-1}$, assuming the positronium line would have a FWHM of 4.2\,MHz, as the positronium line width is thermally broadened to 30 times the hydrogen line width at 1400\,MHz. The positronium upper limit improves upon the results of \cite{Anantharamaiah1989}, who placed a $3\sigma$ upper limit of $<29.3$\,K (recombination rate $<1.1\times10^{44}\,\mathrm{s}^{-1}$, assuming a line width FWMH of 4.2\,MHz) on the detection of the positronium Ps$133\alpha$ line from the Galactic Centre from VLA observations. It should be noted that the recombination rate upper limit from \cite{Anantharamaiah1989} is lower than our value due to the differing flux limits ($<3.4$\,mJy vs. $<84$\,mJy from the VLA and Parkes, respectively) which is a result of the arcsecond vs. arcminute resolution of the VLA ($\sim12\times6$\,arcsec) and Parkes ($\sim15\times15$\,arcmin), respectively.

\begin{table*}
\centering
\begin{tabular}{@{}ccccccc@{}}
\hline
RRL & FWHM [MHz] & $T_{\mathrm{L}}$ [K]$^a$ & $T_{\mathrm{L}}$ [K]$^b$ & $T_{\mathrm{L}}$ [K]$^c$ & $T_{\mathrm{L}}$ [K]$^d$ & $T_{\mathrm{L}}$ [K]$^d$ \\ \hline \hline
\vspace{-8pt} \\
$H165\alpha$ & $0.11\pm0.01$ & $0.30\pm0.02$ & ... & ... & ... & ... \\
$H166\alpha$ & $0.11\pm0.01$ & $0.37\pm0.02$ & 0.65 & 0.14 & $0.09 - 1.03$ & $0.29-0.76$ \\
$H167\alpha$ & $0.13\pm0.01$ & $0.42\pm0.02$ & 0.64 & ... & ... & ... \\
$H168\alpha$ & $0.17\pm0.02$ & $0.54\pm0.04$ & 0.61 & ... & ... & ... \\
$H170\alpha$ & $0.12\pm0.01$ & $0.54\pm0.02$ & ... & ... & ... & ... \\ \hline
Dish  & & 64\,m & 43\,m & 43\,m & 76.2\,m & 25.6\,m \\
Diameter & & & & & & \\  \hline \hline
\end{tabular}
\caption{Galactic Centre FWHM line widths and line temperatures ($T_{\mathrm{L}}$) for hydrogen radio recombination line (RRL) detections from \textit{(a)} this work and $T_{\mathrm{L}}$ from previous studies:  \textit{(b)} \cite{Roberts1970}, \textit{(c)} \cite{Riegel1970}, \textit{(d)} \cite{Kesteven1977}, \textit{(e)} \cite{Hart1980}.}
\label{table:h_recomb}
\end{table*}

\section{CONCLUSIONS}
\label{sec:conclusions}

We have presented results of using a modified ASKAP phased array feed (PAF) mounted on the Parkes 64\,m radio telescope for performing H\,\textsc{i} mapping and stacking. 
\begin{itemize}
\item The standing wave amplitude, resulting from interference with reflected waves, is substantially reduced in the PAF data compared with conventional receivers. We estimate an amplitude reduction by a factor of $\sim 10$ compared with the multibeam receiver. This reduction represents the higher efficiency and full focal-plane sampling of the PAF.

\item The system temperatures during our observations are higher and less stable than those during initial tests by \cite{Chippendale2016}. This is most likely due to delay slips in the digital receiver which were not monitored or corrected for during the observations, but can be corrected using an on-dish noise source to estimate the delays and applying a compensating phase slope to existing beamformer weights.

\item The lower frequency ($\nu<1290$\,MHz) data contain significant but well-known satellite RFI contamination. We demonstrate that, even with the low temporal and spectral resolution of our observations, significant mitigation is possible.

\item We have compared observations of the Large Magellanic Cloud with archival Parkes multibeam data, and found excellent agreement.

\item We have demonstrated that noise continues to decrease with time for long observations with a PAF. In particular, we find a stacked detection of extragalactic H\,\textsc{i} in the GAMA G23 field in the redshift range $0.05 \leq z \leq 0.075$.

\item Two direct H\,\textsc{i} detections in the GAMA G23 field at $z=0.0043$ and $z=0.0055$ of NGC\,7361 and ESO\,469-G015 are also noted. Both integrated spectra show good agreement in spectral shape with archival HIPASS data and the measured fluxes agree within the statistical uncertainties.

\item From targeted observations of HIPASS sources Circinus and NGC\,6744, we found reasonable agreement with the archival HIPASS line of sight spectrum of Circinus. Some of the difference may be due to the different (and not optimal) MPIPAF beam shapes. The integrated line of sight spectrum of NGC\,6744 agrees well with the integrated HIPASS spectrum. 

\item We find clear direct detections of five hydrogen recombination lines: $H165\alpha$, $H166\alpha$, $H167\alpha$, $H168\alpha$ and $H170\alpha$. We do not find a detection of positronium recombination lines in the Galactic Centre observations, but set a $3\sigma$ upper limit of $<0.09$\,K, corresponding to a recombination rate of $<3.0\times10^{45}\,\mathrm{s}^{-1}$.
\end{itemize}

The above demonstration, whilst limited in scope, demonstrates the viability of PAFs on large single dish telescopes. The main areas that need improvement for a permanent installation are the system temperature, which requires cryogenic cooling, and a mechanism for ensuring a stable and reproducible beamforming methodology in the presence of RFI \citep{Chippendale2017}. The flexible beamforming capability of PAFs is nevertheless enormously powerful and can in itself, as already demonstrated, reduce the impact of RFI.

\begin{acknowledgements}
This research was conducted by the Australian Research Council Centre of Excellence for All-sky Astrophysics (CAASTRO), through project number CE110001020. The Parkes radio telescope is part of the Australia Telescope National Facility which is funded by the Commonwealth of Australia for operation as a National Facility managed by CSRIO. The MPIPAF is a collaboration between CSIRO Astronomy and Space Science (CASS) and MPIfR of the Max Planck Society. We wish to thank A. Brown for improving the duty cycle of the 18.5 kHz resolution spectrum integrator, M. Marquarding and E. Troup for contributions to software systems development and integration, particularly for observation and telescope control, and Dr. K. Bannister and C. Haskins for support of software implementation of RFI mitigation. This research has made use of the NASA/IPAC Extragalactic Database (NED) which is operated by the Jet Propulsion Laboratory, California Institute of Technology, under contract with the National Aeronautics and Space Administration.
\end{acknowledgements}

\bibliographystyle{pasa-mnras}
\bibliography{master}

\end{document}